\documentclass{emulateapj} 
\usepackage{apjfonts}

\usepackage{graphics}

\shorttitle{Modeling CAL 83 and CAL 87} \shortauthors{Starrfield
et al.}

\begin{document}
\title{Surface Hydrogen Burning Modeling of Super Soft Xray Binaries:
\\ Are They Supernova Ia Progenitors?}

\author{S. Starrfield\altaffilmark{1}, F. X. Timmes\altaffilmark{2},
W. R. Hix\altaffilmark{3}, E. M. Sion\altaffilmark{4}, W. M.
Sparks\altaffilmark{5}, and S. J. Dwyer\altaffilmark{1}}

\altaffiltext{1}{Department of Physics and Astronomy, Arizona
State University, Tempe, AZ
85287-1504:(sumner.starrfield;steven.dwyer)@asu.edu}
\altaffiltext{2}{T-6, Los Alamos National Laboratory, Los Alamos,
NM, 87545:fxt44@mac.com} \altaffiltext{3}{Department of Physics
and Astronomy, University of Tennessee, Knoxville, TN 37996-1200
\& Physics Division and Joint Institute for Heavy Ion Research,
Oak Ridge National Laboratory, Oak Ridge, TN 37831-6354:
raph@phy.ornl.gov} \altaffiltext{4}{Department of Astronomy,
Villanova University, Villanova, PA: edward.sion@villanova.edu}
\altaffiltext{5}{Science Applications International Corporation,
San Diego CA, 92121 \& X-4, Los Alamos National Laboratory, Los
Alamos, NM, 87545:wms@lanl.gov}

\begin{abstract}
Nova explosions occur on the white dwarf (WD) component of a
Cataclysmic Variable stellar system which is accreting matter lost
by a companion. A Type Ia supernova explosion is thought to result
when a WD, in a similar binary configuration, grows in mass to the
Chandrasekhar Limit. Here, we present calculations of accretion of
Solar matter, at a variety of mass accretion rates, onto hot ($2.3
\times 10^{5}$K), luminous (30L$_\odot$), massive (1.25M$_\odot$,
1.35M$_\odot$) Carbon-Oxygen WDs.  In contrast to our nova
simulations where the WD has a low initial luminosity and a
thermonuclear runaway (TNR) occurs and ejects material, these
simulations do not eject material (or only a small fraction of the
accreted material) and the WD grows in mass.  A hydrogen TNR does
not occur because hydrogen fuses to helium in the surface layers,
and we call this process Surface Hydrogen Burning (SHB). As the
helium layer grows in mass, it gradually fuses either to carbon
and oxygen or to more massive nuclei depending on the WD mass and
mass accretion rate. If such a WD were to explode in a SN Ia
event, therefore, it would show neither hydrogen nor helium in its
spectrum as is observed. Moreover, the luminosities and effective
temperatures of our simulations agree with the observations of
some of the Super Soft X-ray Binary Sources and, therefore, our
results strengthen previous speculation that some of them (CAL 83
and CAL 87 for example) are probably progenitors of SN Ia
explosions. Finally, we have achieved SHB for values of the mass
accretion rate that almost span the observed values of the
Cataclysmic Variables.
\end{abstract}


\section{Introduction}

It is commonly assumed that supernovae of Type Ia (SN Ia) are the
results of thermonuclear runaways (TNR) in the cores of
carbon-oxygen (CO) white dwarfs (WD) which are members of
interacting binary systems, and have accreted material from a
companion until their masses reach the Chandrasekhar Limit (CL)
and a carbon detonation/deflagration occurs (Nomoto, Thielemann,
\& Yokoi 1984; Leibundgut 2000, 2001; Hillebrandt \& Niemeyer
2000).  The binary star systems that evolve to this explosion,
however, have not been identified.  Given the importance of SN Ia,
both to our understanding of the evolution of the Universe and to
the formation of iron peak elements in the Galaxy, we must
identify the progenitors of these explosions.

The binary scenario was originally proposed by Whelan and Iben
(1973), and as a result virtually every type of close binary that
contains a WD has been suggested at one time or another.  However,
based purely on {\it observations}, most of the systems that have
been proposed cannot be the progenitors since the defining
characteristic of a SN Ia outburst is the absence of hydrogen or
helium in the spectrum (Filippenko 1997). (We note, however, that
SN 2002ic, a SN Ia, did show a {\it shell} of hydrogen in its
spectrum [Hamuy et al. 2003].) One of the first suggestions was a
classical nova system (CN), but the amount of core material
ejected during the nova outburst implies strongly that the WD is
decreasing in mass as a result of the outburst (MacDonald 1984;
Gehrz et al. 1998; Starrfield 2003). Other suggestions such as
Symbiotic Novae (T CrB or RS Oph, for example) can probably be
ruled out because the supernova explosion would take place inside
the outer layers of a red giant and hydrogen would be present in
the spectrum (Marietta, Burrows, \& Fryxell 2000; Lentz et al.
2002)

An important development was the discovery of the Super Soft X-ray
Binary Sources (SSS: Tr\"umper et al. 1991) and the suggestions
that they were SN Ia progenitors (van den Heuvel et al. 1992
[VDH92]; Branch et al. 1995; Kahabka \& van den Heuvel 1997
[KVDH97]). They are luminous, L$_{*}\sim 10^{36-38}$ erg s$^{-1}$,
with effective temperatures ranging from 30 to 50 eV or higher ($3
- 7\times 10^5$K). VDH92 proposed that their properties could be
explained by nuclear burning occurring in the outer layers of a WD
at the same rate at which it was being accreted (Steady Burning
Hypothesis: Paczynski and Zytkow [1978)], Sion, Acierno, \&
Tomczyk [1979]; Iben [1982], and Fujimoto [1982a,b]).  Therefore,
the mass of the WD was increasing, it could reach the CL, and then
explode as a SN Ia (VDH92). Nevertheless, except for the
pioneering work of Sion et al. (1979), no stellar evolution
calculations have been done for massive WDs accreting at the high
mass accretion rates required to test the Steady Burning
Hypothesis. Calculations were done for lower mass CO WDs by Iben
(1982) and Sion and Starrfield (1994) but the luminosities and
effective temperatures of their simulations were too low to agree
with the observations of SSS such as CAL 83 or CAL 87 (Greiner
2000), and it was not clear that the WDs that they studied would
reach the CL.

In contrast, we find in the calculations reported here that the
material can burn at mass accretion rates that are far lower than
the Steady Burning rate (a few $\times 10^{-7}$M$_\odot$yr$^{-1}$:
Kahabka 2002).  A hydrogen TNR does not occur because the WD is
sufficiently hot that hydrogen burns in the layers near the
surface as it is accreted. We, therefore, refer to our results as
Surface Hydrogen Burning (SHB) to prevent any confusion with
Steady Burning. In the next section, we briefly describe the SSS.
In the following sections, we describe our evolutionary sequences
and end with a Summary.

\section{Super Soft X-ray Binaries}

The ROSAT all sky survey (and later pointed observations)
established the SSS as a class of X-ray emitting systems
(Tr\"umper et al. 1991).  A recent catalog can be found in Greiner
(2000), and a discussion of their binary properties can be found
in Cowley et al. (1998). Two systems have orbital periods of 1.04
d (CAL 83) and 10.6 h (CAL 87) which are longer than those of
typical CNe. Unfortunately, there is a large range in the
luminosities and effective temperatures reported for a given
system (Greiner 2000). Part of the difficulty is that, for the
systems studied by ROSAT, only a small region of the emitted
spectrum is visible in the detectors and large corrections have to
be applied to the observations (KVDH97).

The tabulated luminosities and effective temperatures of the two
best studied sources are: (1) CAL 83: L = $(0.7 -
6)\times10^{37}$erg s$^{-1}$ and T$_{\rm eff}$ = $(5 - 6) \times
10^5$K (Parmar et al. 1998). The analyses of XMM Reflection
Grating Spectrometers spectra of CAL 83 imply a temperature $\sim
5 \times 10^5$K (Paerels et al. 2001). (2) CAL 87: L = $(3 -
5)\times10^{36}$erg s$^{-1}$ and T$_{\rm eff}$ = $(6 - 9) \times
10^5$K (Parmar et al. 1997). The WD masses are, as yet,
undetermined but the temperatures and luminosities imply WD radii.
Observed properties for a number of SSS (values taken from the
above references or Greiner 2000) are shown in a theoretical HR
diagram (Figure 1) along with the results of our calculations
(described below). The systems shown in this figure are CAL 83,
CAL 87, RX J0019.8+2156 (RXJ0019), RX J0925.7-4758 (RXJ0925), 1E
0035.4-7230 (1E0035), and RX J0439.8-6809 (RXJ0439). RXJ0019 and
RXJ0925 are in our Galaxy, CAL 83, CAL 87, and RXJ0439 are in the
LMC, and 1E0035 is in the SMC.

While CAL 87 appears to be less luminous than CAL 83, it is an
eclipsing binary with an eclipse depth of 2 mag (Cowley et al.
1990). An XMM observation suggests that we are seeing only the
accretion disk in X-rays and the WD must be more luminous (Orio et
al. 2004).  A factor of two increase in luminosity would place it
on top of our results for 1.35M$_\odot$. The WD is probably this
massive (Cowley et al. 1998).

\section{Surface Hydrogen Burning Sequences for SSS}

We use the one-dimensional, fully implicit, Lagrangian,
hydrodynamic computer code described in Starrfield et al. (1998,
2000) with one major change.  The nuclear reaction network used in
those papers was that of Weiss and Truran (1990).  In this work,
we use the pp+cno+rp network of Timmes \footnote{\tt
www.cococubed.com/code\_pages/net\_pphotcno.shtml}.  We also use
the equation of state of Timmes (Timmes and Arnett 1999; Timmes
and Swesty 2000) and the OPAL opacities (Rogers, Swenson, \&
Iglesias 1996).   We include the accretion energy as described in
Shaviv and Starrfield (1988) although it is far smaller than the
nuclear energy.

Our initial model is obtained by evolving a WD through a CN
outburst. Once {\it all} the ejected material is expanding faster
than the escape speed, has reached radii exceeding 10$^{13}$cm,
and is optically thin; we remove it from the calculations.  We
then take the remnant WD and allow it to cool to $\sim
10^{-2}$L$_\odot$ before restarting accretion. We do this for
three nova cycles, take the remnant WD from the last cycle, and
then begin accretion when the WD luminosity has decreased to
$\sim$30 L$_\odot$.  (This luminosity is that found for V1974 Cyg
three years after outburst: Shore et al. 1997).  We rezone our
initial model so there are 95 zones covering the entire WD (our
results are independent of the number of zones but not the zone
mass: see below).  The initial model has an outer boundary mass of
$10^{-22}$M$_\odot$, a surface zone mass of $10^{-5}$M$_\odot$,
and a temperature exceeding $7 \times 10^7$K. (We report the
effects of the initial luminosity on the evolution in a later
paper.)   We used this procedure to calculate initial models for
1.25M$_\odot$, and 1.35M$_\odot$ WDs.  The core composition was
50\%\ carbon and 50\%\ oxygen.  Solar composition material was
accreted onto these WDs at a range of mass accretion rates. The
metallicity of the LMC is about one-third that of the Solar
neighborhood so we will study the effects of metallicity in a
future paper.  This will be important since metallicity clearly
affects the evolution of SN Ia (Timmes et al. 2003), and the range
of $z$ over which SN Ia are currently being studied implies that
they have different metallicities.

We begin the evolution with a bare CO core.  In Sequence 4, for
example, we find that with a surface zone temperature of $7 \times
10^7$K and density of $10^4$ gm cm$^{-3}$, the energy generation
($\epsilon_{nuc}$) rises quickly to 10$^{4}$erg gm$^{-1}$s$^{-1}$.
The outer layers are strongly heated by this energy release and,
after 15 yr of evolution, T$_{\rm sz}$ (the temperature in the
surface zone) = $2.7 \times 10^8$K, $\epsilon_{nuc}$ = $3.5 \times
10^{9}$erg gm$^{-1}$s$^{-1}$, and L$_{SHB}$ = $1.3 \times 10^{38}$
erg s$^{-1}$.  It is now sufficiently hot in the surface zone so
that it takes less time than the time step ($\sim 2 \times 10^6$s)
for all the infalling hydrogen to burn to helium in this zone. In
addition, some of the helium is already burning to carbon in this
zone.  We designate this process Surface Hydrogen Burning (SHB).
Similar behavior, hydrogen burning out in the surface zone, occurs
in all sequences listed in Table 1.  The helium mass fraction
declines to zero somewhat deeper into the WD and would have too
small an abundance to be seen in the spectrum if this WD were to
explode.

We find SHB at both WD masses but, for brevity, present only part
of the results for 1.35M$_\odot$ in Table 1 and Figure 1. The
results for other masses, and a detailed discussion of all our
evolutionary sequences, will appear elsewhere (Starrfield et al.
2004). The rows in Table 1 are the mass accretion rate
(M$_{\odot}$ yr$^{-1}$), the length of time in years that we have
followed the evolution, the amount of mass accreted (M$_\odot$),
the temperature (T$_{\rm sz}$) and rate of energy generation
($\epsilon_{\rm sz}$) in the surface zone (sz), the luminosity,
and the effective temperature (both in K and eV) of the
evolutionary sequence. They are tabulated at the time we end the
evolution. We stopped the evolution in Sequences 3, 4, and 5 when
the central density approached values where carbon burning, not
included in our network, should have begun. Sequence 1 exhibited a
TNR in the helium layer after $\sim 5\times 10^4$ yr of evolution
but insufficient energy was released to directly eject any
material. This sequence achieved a peak temperature in the TNR
exceeding $7 \times 10^8$K and will be redone with a larger
network. The other evolutionary sequences were hotter and not as
degenerate when helium burning began so that no flash occurred.
The final rows give the composition at the interface between the
accreted matter and the core matter (CI) near or at the end of
evolution.

Our results show that SHB occurs at 1.35M$_\odot$ for values of
the mass accretion rate typically observed in Cataclysmic
Variables (CVs). This implies that if a post-CN system, with a
massive WD, can arrive at a configuration where mass transfer onto
the WD begins when the WD is still hot and luminous, then the mass
of the WD will grow {\it even if it is accreting at a rate
normally observed only for CVs}. The system does not have to be
accreting {\it at} or near to the canonical Steady Burning rate
for the WD to grow in mass to the CL.

Our most important result is that in none of our sequences is
there any hydrogen (and little helium) left in the surface mass
zones on the WD. Additional sequences done with finer mass zoning
at the surface ($10^{-7}$M$_\odot$ or 100 times smaller than the
sequences in this Letter) show that hydrogen and helium reach only
to a depth of $\sim 10^{-6}$M$_\odot$.  This implies that if such
a structure were to explode in a SN Ia outburst, then there would
be insufficient hydrogen or helium to appear in the spectrum
(Lentz et al. 2002).

We also find that for sequences 2, 3, and 4, as the mass accretion
rate increases, the abundance of $^{12}$C decreases and that of
$^{16}$O increases. In Sequence 1 the helium TNR was responsible
for fusing the nuclei to the top of the included network (mass 20)
and Sequence 5 was sufficiently hot to burn almost all the nuclei
to this mass. Therefore, we predict that the C/O ratio in the
topmost layers of the ejecta (observed first and moving the
fastest) will vary between Ia explosions and, when measured, would
provide an estimate of the accretion rate prior to the Ia
explosion.

The accretion rate in Sequence 5 is above the assumed Steady
Burning rate but it still undergoes SHB and does not grow rapidly
to large radii and shut off accretion.  A sequence was evolved
with an accretion rate of $1.6\times 10^{-6}$M$_{\odot}$ yr$^{-1}$
and it expanded to large radii after 30 years of evolution.

Much larger amounts of mass were accreted onto these WDs than in
our previous studies of accretion onto low luminosity WDs done for
studies of CN outbursts (Starrfield et al. 2000, and references
therein).  In one case (Sequence 5), the mass of the WD increased
from 1.35M$_\odot$ to $\sim$1.38M$_\odot$ and the central density
grew to $2 \times 10^9$g cm$^{-3}$.  Clearly, these calculations
need to be extended with carbon burning reactions included in the
nuclear reaction network.

We compare the results of our calculations to the observations in
Figure 1 which plots both the Luminosity (L$_{\rm SHB}$) and
effective temperature of the sequences together with the
observations in a theoretical HR-Diagram. The final value for each
of our sequences is plotted as an asterisk (*) and they are
connected by a line to guide the eye. The observed ranges for the
SSS are plotted as boxes. CAL83 and 1E0035.4 fall close to the
results for 1.25M$_\odot$, while CAL 87 and RXJ0925 fall close to
the results for 1.35M$_\odot$. We predict that these four systems
probably contain massive, hot WDs accreting at high rates and
growing to the CL. While CAL 87 falls below the line for
1.35M$_\odot$, because it is eclipsing the WD is probably more
luminous. The other two SSS are more problematic. RXJ0019 and
RXJ0439 are in a region of the HR diagram suggesting that they
either contain lower mass WDs or are recovering from an extended
traverse to cooler effective temperatures.  Given the
uncertainties in the observations and the fact that we have only
begun this study, the agreement is encouraging.

\begin{figure}
\includegraphics[angle=90,scale=.40]{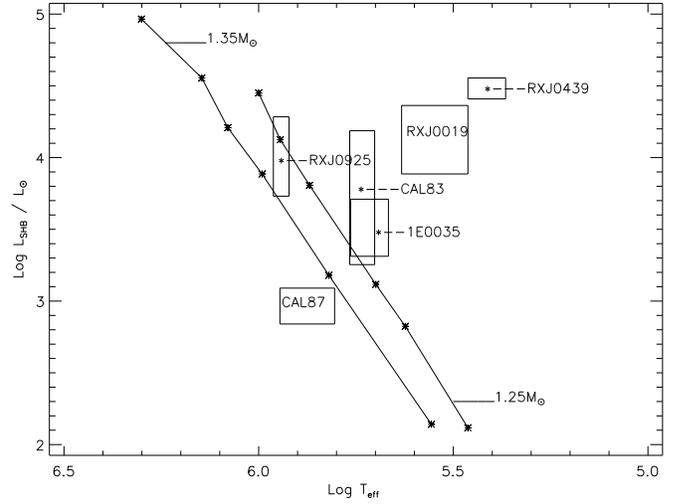}
\caption{We compare the luminosities (L$_{\rm SHB}$) and effective
temperatures of our sequences for 1.35M$_\odot$ and 1.25M$_\odot$
to the observed X-ray luminosities and temperatures for well
studied SSS.  Because the observers only give ranges, we plot
their data as boxes with a label indicating the system.  The final
values for each of our sequences are plotted as an asterisk (*)
connected by a line to guide the eye. The line is labeled by the
WD mass.}
\end{figure}

\section{Summary}

We have investigated the effects of accretion at high rates onto
luminous (30L$_\odot$), hot (T$_{\rm eff}$=$2.3 \times 10^{5}$K),
and massive (1.25M$_\odot$ and 1.35M$_\odot$) WDs.  We find
quiescent burning of hydrogen and helium in the surface layers
and, thereby, the growth of the WD to the Chandrasekhar Limit.
Quiescent burning occurs at a large range of mass accretion rates
and the luminosities and effective temperatures of our
evolutionary sequences agree with those observed for the SSS which
makes it probable that systems such as CAL 87 and CAL 83 are the
progenitors of SN Ia.  We also find, for all but the lowest mass
accretion rate, that all the accreted hydrogen and helium is
transformed to at least carbon and oxygen in the surface mass
zones.  If one of our evolutionary sequences were to explode as a
SN Ia, then there would be less than $\sim 10^{-6}$M$_\odot$ of
hydrogen and helium in the expanding layers which is unobservable
(Lentz et al. 2002).   We also find that the ratio of C/O in the
surface zones is a function of mass accretion rate and should vary
from one SN Ia explosion to another.

An exciting implication of this study is that a Cataclysmic
Variable binary system can produce either a CN or a SSS depending
primarily on the luminosity of the WD when accretion is initiated
after a CN outburst. Therefore, we propose first, that a CN and a
SSS may be two different phases of the evolution of the same
binary system and, second, that a CV system may evolve from a SSS
to a CN or the reverse at different times of its evolution.

We also find:

\noindent{\bf 1)} If CAL 83 is a 1.25M$_\odot$ WD, then it is
accreting at a rate between $2 \times 10^{-8}$ M$_{\odot}$
yr$^{-1}$ and $8 \times 10^{-8}$ M$_{\odot}$ yr$^{-1}$. This is
less than the Steady Burning mass accretion rate. {\bf 2)} If CAL
87 is a 1.35 M$_\odot$ WD, then it is also accreting at a rate
between $2 \times 10^{-8}$ and $8 \times 10^{-8}$ M$_{\odot}$
yr$^{-1}$.  Since CAL 87 is eclipsing, we assume that we are
observing only a fraction of the energy emitted by the WD. {\bf
3)} Our initial models were computed by accreting onto WDs that
had just experienced a CN explosion but had not yet cooled to low
luminosities.  If this evolution is realized in nature, then we
predict a link between CN and SSS which results in SN Ia
explosions. {\bf 4)} Sequences 3, 4, and 5 have accreted a great
deal more material than found in studies of accretion onto cool
WDs. We feel confident that if we were to continue the evolution
for the necessary time, then they would reach the Chandrasekhar
Limit. {\it Thus, they satisfy the conditions necessary to be
considered as strongly viable candidates for the progenitors of SN
Ia explosions}

\noindent {\it Acknowledgements} We are grateful to E. Baron, A.
Cowley, A. Filippenko, S. Kenyon, J. Krautter, J. MacDonald, A.
Mezzacappa, S. Rappaport, P. Schmidtke, and J. Truran for valuable
discussions. We are grateful to the referee for comments which
improved this Letter. S. Starrfield acknowledges partial support
from NSF and NASA grants to ASU. He also thanks J. Aufdenberg and
ORNL for generous allotments of computer time. FXT is supported by
the DOE under Grant No.~B341495 to the Center for Astrophysical
Thermonuclear Flashes at the University of Chicago and by the
National Security Fellow program at Los Alamos National
Laboratory. WRH is partly supported by the NSF under contracts
PHY-0244783 and and by the DOE, through the Scientific Discovery
through Advanced Computing Program. ORNL is managed by
UT-Battelle, LLC, for the U.S. DOE under contract
DE-AC05-00OR22725.

\begin{deluxetable}{@{}lccccc}
\tablecaption{1.35M$_\odot$ Hot White Dwarf Evolutionary Sequences
\tablenotemark{a}} \tablewidth{0pt} \tablehead{ \colhead{Sequence}
& \colhead{1\tablenotemark{b}}   & \colhead{2} & \colhead{3} &
\colhead {4} & \colhead{5}  } \startdata \.M (M$_{\odot}$
yr$^{-1}$)&$1.6\times10^{-9}$
&$1.6\times10^{-8}$&$1.6\times10^{-7}$&$3.5\times10^{-7}$&$8.0\times10^{-7}$ \\
$\tau_{\rm evol}$ (yr)&$8.3\times10^5$
&$2.2\times10^5$&$1.3\times10^5$&$3.8\times10^4$&$4.8\times10^4$ \\
$\delta$M$_{\rm
acc}$(M$_\odot$)&$1.3\times10^{-3}$&$3.5\times10^{-3}$&
$2.1\times10^{-2}$&$1.3\times10^{-2}$&$3.8\times10^{-2}$ \\
T$_{\rm sz}$($10^6$K)\tablenotemark{c}&114&177&319&347&516 \\
$\epsilon_{\rm sz}$
(10$^{8}$erg gm$^{-1}$s$^{-1}$)&0.15 &1.5&16.6&36.8&91.0  \\
L$_{\rm SHB}$ (erg
s$^{-1}$)&$5.4\times10^{35}$&$5.9\times10^{36}$& $6.3\times
10^{37}$ & $1.4\times 10^{38}$& $3.6\times 10^{38}$   \\
T$_{\rm eff}$(K)&$3.6\times10^5$&$6.6\times10^5$&
$1.2\times10^6$&$1.4\times10^6$&$2.0\times10^6$ \\
T$_{\rm eff}$(eV)&32&57&107&125&175 \\
$^1$H(CI)\tablenotemark{d}&0.0&0.0&0.0&0.0&0.0 \\
$^4$He(CI)&$<$0.01&0.0&0.0&0.0&0.0 \\
$^{12}$C(CI)&0.11&0.45& 0.31&0.22&0.06  \\
$^{13}$C (CI)&0.01&0.05&0.13&0.14&0.12 \\
$^{14}$N (CI)&0.02&0.35&0.13&0.10&0.06 \\
$^{16}$O (CI)&0.01&0.15&0.42&0.49&0.07 \\
A$>$19 (CI)&0.77&$<$0.01& $<$0.01&$<$0.01&0.60 \\
\enddata
\tablenotetext{a}{The initial model for all evolutionary sequences
had M$_{\rm WD}$=1.35M$_\odot$, L$_{\rm WD}$=30L$_\odot$, T$_{\rm
eff}$=$2.3 \times 10^{5}$K, and R$_{\rm WD}$=2391 km}
\tablenotetext{b} {Sequence 1 experienced a helium TNR in which
the temperature exceeded $7\times10^8$K} \tablenotetext{c} {sz =
surface zone} \tablenotetext{d}{ CI= Composition Interface: All
abundances are mass fractions.}
\end{deluxetable}

\vfill\eject


\begin{thebibliography}{99}



\bibitem{branch95} Branch, D., Livio, M., Yungelson, L. R., Boffi,
F. R., Baron, E. 1995, PASP, 107, 1019

\bibitem{cowl90} Cowley, A., Schmidtke, P.,  Crampton, D., Hutchings,
J. 1990, ApJ, 350, 288

\bibitem{cowl98} Cowley, A., Schmidtke, P.,  Crampton, D., Hutchings, J. 1998, ApJ, 504,
854

\bibitem{fil97} Filippenko, A. V. 1997, ARAA, 35, 309

\bibitem{Fuj82a} Fujimoto, M. Y. 1982a, ApJ, 257, 752

\bibitem{Fuj82b} Fujimoto, M. Y. 1982b, ApJ, 257, 767

\bibitem{gehrz98} Gehrz, R.D., Truran, J.W., Williams, R.E., \&
Starrfield, S. 1998, PASP, 110, 3.


\bibitem{grein96} Greiner, J. 2000, New Astronomy, 5, 137

\bibitem{Hamuy03} Hamuy, M., et al. 2003, NATURE, 424, 651

\bibitem{hill2000} Hillebrandt, W., \& Niemeyer, J. 2000, ARAA,
38, 191

\bibitem{Iben82} Iben, I. 1982, ApJ, 259, 244

\bibitem{kah02} Kahabka, P. 2002, in Compact Stellar X-ray
Sources, ed. W. Lewin \& M. van der Klis, Cambridge University
Press (astro-ph:0212037)


\bibitem{kah} Kahabka, P., van den Heuvel, E. P. J. 1997, ARAA, 35, 69


\bibitem{leib00} Leibundgut, B. 2000, A\&A Reviews, 10, 179

\bibitem{leib01} Leibundgut, B. 2001, ARAA, 39, 67

\bibitem{lentz02} Lentz, E., Baron, E., Hauschildt, P. H., \&
Branch, D. 2002, ApJ, 580, 274

\bibitem{macd84} MacDonald, J. 1984, ApJ, 283, 241

\bibitem{marburfry00} Marietta, E., Burrows, A., \& Fryxell, B.
2000, ApJS, 128, 615

\bibitem{nom82} Nomoto, K., Thielemann, F.-K., \& Yokoi, K. 1984,
ApJ, 286, 644

\bibitem{orio04} Orio, M., Ebisawa, K., Heise, J., \& Hartmann, W.
2004, in Compact Binaries in the Galaxy and Beyond, in press
(astro-ph/0402040)

\bibitem{paczzytk78} Paczynski, B., \& Zytkow, A. N. 1978, ApJ,
222, 604

\bibitem{pae01} Paerels, F., Rasmussen, A. P., Hartmann, H. W., Heise, J.,
Brinkman, A. C., de Vries, C. P., den Herder, J. W. 2001, A\&A, 365, L308

\bibitem{parm97} Parmer, A. N., Kahabka, P., \& Hartmann, H. W. 1997, A\&A, 323, L33

\bibitem{parm98} Parmer, A. N., Kahabka, P., Hartmann, H. W.,
Heise, J., \& Taylor, B. G. 1998, A\&A, 332, 199



\bibitem{opal96} Rogers, F. J., Swenson,  Iglesias, C. A. 1996,
ApJ, 456, 902

\bibitem{ShaSta88} Shaviv, G., \& Starrfield, S. 1988, ApJ, 335,
383

\bibitem{shore97} Shore, S. N., Starrfield, S., Ake, T., \&
Hauschildt, P. H. 1997, ApJ, 490, 393.

\bibitem{sionetal79} Sion, E. M., Acierno, M. J., \& Tomczyk, S.
1979, ApJ, 230, 832

\bibitem{siosta94} Sion, E. M., \& Starrfield, S. 1994, ApJ, 421,
261

\bibitem{starr03} Starrfield, S. 2003, in From Twilight to
Highlight: The Physics of Supernovae, ed. W. Hillebrandt \& B.
Leibundgut, Springer, Heidelberg, p. 128

\bibitem{starr00} Starrfield, S., Sparks, W. M., Truran, J.W., Wiescher, M.C. 2000,
ApJS, 127, 485

\bibitem{starr04} Starrfield, S., Timmes, F. X., Hix, W. R., Sion,
E. M., Sparks, W. M., Dwyer, S. J. 2004, in preparation.

\bibitem{starr98} Starrfield, S., Truran, J.W., Wiescher, M.C., \&
Sparks, W. M. 1998, MNRAS, 296, 502


\bibitem{TandA99} Timmes, F. X., \& Arnett, D. A. 1999, ApJS, 125,
277

\bibitem{timbrotru03} Timmes, F. X., Brown, E. F., Truran, J. W.
2003, ApJ, 590, L83


\bibitem{TandS00} Timmes, F. X., \& Swesty, D. 2000, ApJS, 126,
501

\bibitem{trump91} Tr\"umper, J., Hasinger, G., Aschenbach, B.,
Br\"auninger, H., Briel, E. G. et al. 1991, NATURE, 349, 579

\bibitem{vandenh} van den Heuvel, E. P. J., Bhattacharya, D.,  Nomoto, K., Rappaport, S. A.
1992, A\&A, 262, 97

\bibitem{weitru90} Weiss, A., \& Truran, J. W. 1990, A\&A, 238,
178

\bibitem{Wheliben} Whelan, J., Iben, I. 1973, ApJ, 186, 1007



\end{thebibliography}
\end{document}